\documentstyle[prl,aps,tighten,multicol,epsfig]{revtex}
\begin{document}
\title{Beyond the Zipf-Mandelbrot law in quantitative linguistics}
\author{Marcelo A. Montemurro}
\address{Facultad    de    Matem\'atica,    Astronom\'{\i}a    y    F\'{\i}sica,
Universidad Nacional de C\'ordoba, \\          Ciudad Universitaria, 5000
C\'ordoba, Argentina}
\maketitle

\begin{abstract}
In  this  paper  the  Zipf-Mandelbrot   law  is  revisited  in  the  context  of
linguistics. Despite its widespread popularity the Zipf--Mandelbrot law can only
describe the statistical behaviour of a rather restricted fraction of the  total
number of  words contained  in some  given corpus.  In particular,  we focus our
attention on   the important deviations that  become statistically  relevant  as
larger corpora are considered and that ultimately could be understood as salient
features of the underlying  complex process of language generation.  Finally, it
is shown that all the  different observed regimes can be  accurately encompassed
within a single mathematical framework recently introduced by C. Tsallis.
{{\bf Pacs Numbers}: 89.75.Fb--Structure and organization in complex systems  \and
      89.90.+n--Other topics in areas of applied and interdisciplinary Physics (restricted to
new topics in section 89)
}  \\
\end{abstract}

\begin{multicols}{2}

\section{Introduction}

In 1932 George Zipf \cite{zipf1} put forward an empirical observation on certain
statistical regularities of human writings  that has become the most  well known
statement of  quantitative linguistics.  He found  that in  a given  text corpus
there  is  an  approximate  mathematical  relation  between  the  frequency   of
occurrence of each word and  its rank in the list  of all the words used  in the
text ordered by decreasing frequency. He also pointed out similar relations that
hold in  other contexts  as well  \cite{zipf1,zipf2}, however,  in this work  we
shall just concentrate  on its applications
in linguistics.

Let us identify  a particular word  by an index  $s$ equal to  its rank, and  by
$f(s)$ the normalised frequency of occurrence of that word, that is, the  number
of times it appears in the text divided by the total number of words $N$.  Then,
Zipf's law states that the following relation
holds approximately:
\begin{equation}
\label{zipf} f(s)=
\frac{A}{s^{\alpha}} \quad ,
\end{equation}
where the exponent $\alpha$ takes on a value slightly greater than $1$, and  $A$
is a normalising constant. Although this is a strong quantitative statement with
ubiquitous applicability  attested over  a vast  repertoire of  human languages,
some observations are  in place. First,  Zipf's law in  its original form  as we
have written  it, can  at most  account for  the statistical  behaviour of words
frequencies in a rather limited  zone in the middle-low to low range of the rank
variable. Even in the case of long single texts Zipf's law renders an acceptable
fit in the small window between $s\approx 100$ and $s \approx 2000$, which  does
not represent  a significant  fraction of  any literary  vocabulary. Second, the
modification introduced  by Mandelbrot \cite{mandel1} by using  arguments on the
fractal  structure  of lexical  trees,  though valuable  in  terms of   possible
insights  into  the  statistical manifestations  derived  from  the hierarchical
structure of languages,  has not a  notorious impact on  quantitative agreements
with empirical data. In fact, the only improvement over the original form of the
law is  that it  fits more  adequately the  region corresponding  to  the lowest
ranks,  that  is $s<100$,  dominated  by function  words.  The generalised  form
proposed by Mandelbrot can be written as follows:
\begin{equation}
\label{mandel2}
f(s)=\frac{A}{(1+Cs)^\alpha} \quad ,
\end{equation}
where $C$ is a second parameter that needs to be adjusted  to fit the data.

It has been shown that this form  of the law is also obeyed by  random processes
that can be mapped onto texts \cite{mandel2,li}, hence ruling out any sufficient
character   for  linguistic   depth  inherent   to  the   Zipf-Mandelbrot  law.
Nevertheless, it  has been  argued that  it is  possible to discriminate between
human  writings  and  stochastic  versions  of  texts  precisely  by  looking at
statistical  properties  of words  that  fall beyond  the  scope where  equation
(\ref{mandel2}) holds \cite{havlin}.

In this paper two complementary goals  are pursued. In the first place  we  wish
to display evidence  on the statistical  significance of the  {\em non  Zipfian}
behaviour that emerges as a robust feature  when large corpora are   considered.
Second, we shall present a complete mathematical framework by which we  redefine
Zipf's  law  in   order  to  describe   in  a  precise   manner  the   empirical
frequency~-~rank distributions of words in  human writings over the whole  range
of the rank variable. We believe that an accurate phenomenological understanding
can  throw light  in  the   search  of plausible  microscopic  models   as  well
as   dictate  necessary features  that  those  models  should  eventually comply
with. Moreover, the variety of  mechanisms that have  been proposed to   explain
Zipf-Mandelbrot  law  \cite{mandel2,li,simon,pietro}     do  not   predict   any
anomaly associated to high-rank words. Whence, the genuine statistical  features
that we will address in our analysis  may lead to  a more reliable  hallmark  of
complexity  in human writings.

\section{Statistical evidence for a reformulation of Zipf's law}

Generally,  Zipf's  plots obtained  for  single texts  suffer  from a  lack  of
sufficient statistics in  the region corresponding  to high values  of the rank
variable, that  is, for  words that  appear just  a few  number of times in the
whole text.  Possibly, that  may have  been one  of the  reasons why  so little
attention has been devoted so far to the distribution  of words from $s \approx
2000$ onwards. To resolve  the behaviour of those  words we need a  significant
increase in volume  of data, probably  exceeding the length  of any conceivable
single text.  Still, at  the same  time it  is desirable  to maintain as high a
degree of homogeneity  in  the texts as  possible,  in the hope of  revealing a
more complex phenomenology than that simply originating from a bulk average  of
a wide range of disparate sources.

In the Zipf's graphs we show in this paper, the presented data points correspond
to averages over non  overlapping windows on the  rank variable, centred at  the
displayed points. The windows' widths are constant in the logarithmic scale  and
this average is done in order to smooth local fluctuations in the data.

\begin{figure}
\label{figure1}
\epsfig{file=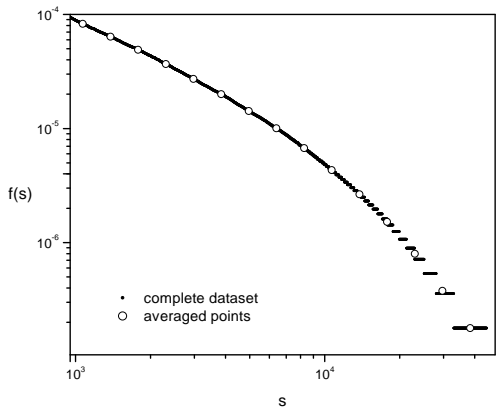,width=83mm}
\caption{Last section  of a  Zipf plot  where we  show the  actual data  and the
averaged values we  use to represent  the data. The  region of highest  ranks is
dominated by the  long plateaux associated  with very infrequent  words. Usually
the last one or two points ought to be discarded. The data correspond to a collection
of 56 books by Charles Dickens.}
\end{figure}

Figure~1 shows  the last  part of  a typical  Zipf plot  for a large text corpus
depicting the  actual data  together with  the averaged  values. The  step--like
plateaux are a finite size effect due to words that appear just a few number  of
times and are indeed a consequence of poor statistics for the highest values  of
the  rank.   The  last  one  of  these  plateaux,  which  is  also  the longest,
corresponds   to hapaxes,  that is  words that  appear just  once in  the text.
Therefore, in  any quantitative  discussion on  the form  of the  frequency-rank
distribution some of the last points, usually one or two, should be  disregarded
on the basis of the foregoing observations.

Along these lines, in Figure~2 we show the Zipf's graphs obtained for four large
corpora, each gathering several works from four different authors  respectively.
In the figures, $N$ represents the total number of tokens present in the
corpus, and $V$ the vocabulary size.
It  is only  when we  start to  analyse large  samples like  these that   robust
statistical  features begin  to emerge  in the  region belonging  to the  higher
ranks. The most conspicuous observation is  that all the curves start to  depart
from the power--law regime at  approximately $s\approx 2000-3000$, and then  all
have a tendency to a faster decay that is slightly different for each curve. All
the curves roughly agree in the region where Zip's law holds and each again  has
a different behaviour in the lowest ranks. In total three
regimes are clearly distinguished in the four data sets.

It is also interesting to note that despite the vocabularies and styles vary  in
the  four  corpora  considered,  the  natural  divisions  in  the  qualitatively
different regimes look very similar. More specifically, whereas the intervals on
the rank axis that cover the first two regimes are roughly of the same size  for
all  the four  corpora, the  difference in  vocabulary length  reflects on  the
differences  in the  length of  the fast-decaying  tails for  each corpus.  This
suggests that  regardless of  the different  sizes of  the texts considered, the
vocabularies   can   be   divided   into   two   parts   of   distinct    nature
\cite{cancho1,cancho2}: one  of basic  usage whose  overall linguistic structure
leads to  the Zipf-Mandelbrot  law, and  a second  part containing more specific
words with a less  flexible syntactic function.

\begin{figure}
\label{figure2}
\epsfig{file=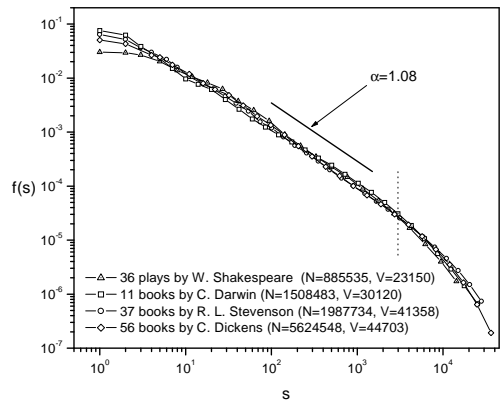,width=83mm}
\caption{Frequency-rank distribution of  words for four  large text samples.  In
order to  reveal individual  variations these  corpora are  built with  literary
works of four different authors respectively. The vertical
dash line is placed approximately where Zip's law ceases to hold.}
\end{figure}

The question now may arise as to whether there is a kind of asymptotic behaviour
as even larger corpora  are analysed. It is  clear that in order  to answer this
we are compelled to release a certain degree the constraints on homogeneity  and
consider  samples from  various authors  and styles.  In Figure~3  we show  the
frequency-rank distribution of words  in a very large  corpus made up of  $2606$
books written in English comprising nearly 1.2GB of ASCII data. The total number
of tokens in this  case rose to $183403300$  with a vocabulary size  of $448359$
different words. It  is remarkable that  the point at  which the departure  from
Zipf's law takes place has just  moved to $r \approx 6000$ despite  the increase
in  sample size.  However, the  striking new  feature is  that the  form of  the
distribution for  high ranks  reveals as  a second  power law  regime. This last
result is  an independent  confirmation of  a similar  phenomenology observed by
Ferrer and Sol\'e in a very large corpus  made up of a collection of samples  of
modern  English, both  written and  spoken, each  no  longer  than 45000  words
\cite{cancho1}.  There,  they  found  that  the  second  power  law  regime  was
characterised by  a decay  exponent close  to $-2$.  However,  a  precise direct
measurement of this decay exponent poses some difficulty since the last  portion
of  the  Zipf  plot may  still  be   affected by  the  poor  statistics of  very
infrequent  words.  That may  translate  into a  slight  underestimation of  the
absolute value of the  exponent. Despite this caveat,  in Figure~3 we have  also
shown  a pure  power law with  decay exponent $\alpha=2.3$,  solely as a  visual
reference.

\begin{figure}
\label{figure3}
\epsfig{file=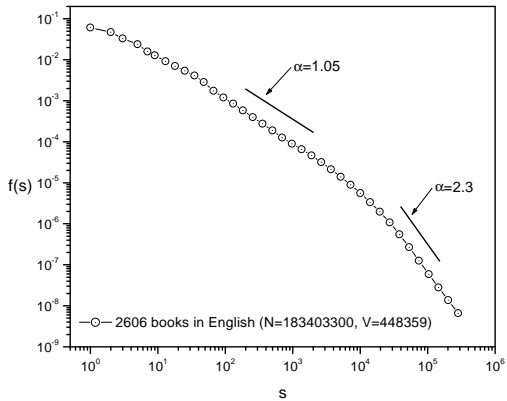,width=83mm}
\caption{Zipf's plot for a large corpus comprising 2606 books in English, mostly
literary works  and some  essays. The   straight lines  in the logarithmic graph
show pure power laws as a visual aid. The last point of  the in the Zipf's  plot
was eliminated since it is severely affected by the plateaux associated with the
least frequent words}
\end{figure}

The  main purpose  of this  section was  to show  that even  though there  is a
restricted domain in which the universality of Zipf's law seems to be valid, new
and solid statistical regularities emerge as the sample size is increased. These
regularities are no less impressive than the original observations made by G. K.
Zipf, and, furthermore, they might be more deeply related to particular features
of the complex process of language generation.

In the next section  we will present a  mathematical model within which  all the
phenomenology discussed here can be quantitatively described.

\section{The mathematical model}

We start from the simple observation that the Zipf-Mandelbrot law satisfies  the
following  first  order differential  equation,  as can  be  verified by  direct
substitution:
\begin{equation}
\label{mand2}
\frac{d f}{d s}=-\lambda f^q \quad .
\end{equation}
The solutions  to equation  (\ref{mand2}) asymptotically  take the  form of pure
power  laws  with decay  exponent  $1/(q-1)$. It  is  possible to  modify   this
expression  in  order  to include  a  crossover  to another  regime,  as  in the
following  more general equation:
\begin{equation}
\label{diff}
\frac{d f}{d s}=-\mu f^r-(\lambda-\mu)f^q \quad  .
\end{equation}
where  we have added a new parameter  and a new exponent. In the case  $1\le r<q$
and $\mu \neq 0$ the effect of the new additions is to allow the presence of two
global regimes characterised  by the dominance  of either exponent  depending on
the  particular  value  of  $f$.  The use  of  this  equation  in  the realm  of
linguistics  was originally suggested by C. Tsallis \cite{tsallis1}, and it  had
previously been  used to  describe experimental  data on  the re-association  in
folded   proteins   \cite{tsallis2} within   the   framework   of  non--extensive
Statistical  Mechanics \cite{tsallis3}.  It is  worth mentioning  here that  the
Zipf-Mandelbrot  law for  words, equation (\ref{mandel2}), has  been  related
to  Tsallis' generalised Thermodynamics by means of heuristic arguments based on
the fractal structure of symbolic sequences with long-range correlations
\cite{denisov}.

Some  qualitative features  of the  solutions of  equation (\ref{diff})  can be
grasped by  analysing different  possibilities for  the involved  parameters. We
summarise those in the following three representative cases and refer the reader
to reference \cite{tsallis2} for a more complete discussion:

\begin{itemize}
\item Let  us note  first that  by taking  $r=q>1$, or  equivalently $\mu=0$ and
$q>1$, we recover equation (\ref{mand2}) the usual form of Zipf-Mandelbrot  law,
since by direct integration we have
\begin{equation}
\label{tsallis1}
f(s)=\frac{1}{\bigl [1+(q-1)s\lambda \bigr ]^{\frac{1}{q-1}}} \quad ,
\end{equation}
where we chose $f(0)=1$. \item Now, letting $r=1$ and taking $q>1$ one obtains
\begin{equation}
\label{tsallis2}
f(s)=\frac{1}{\bigl[1-\frac{\lambda}{\mu}+\frac{\lambda}{\mu} e^{(q-1)\mu s}
\bigr ]^{\frac{1}{q-1}}} \quad .
\end{equation}
This expression shows  a very interesting  behaviour for $\mu<<\lambda$,  since
for small values of  $s$ it reduces to  equation (\ref{tsallis1}) and then  for
larger values of $s$ it undergoes a crossover to an exponential decay.
\item Finally let us consider the more general situation $1<r<q$.
In this case the integration yields
 \begin{eqnarray}
\label{tsallis3}
s&=&\frac{1}{\mu} \Bigl \{
\frac{f^{-(r-1)}-1}{r-1}-\frac{(\lambda/\mu)-1}{1+q-2r} \nonumber \\ 
&\times&\bigl[H(1;q-2 r,q-r,(\lambda/\mu)-1) \nonumber \\
&-& H(f ;q-2r,q-r,(\lambda/\mu)-1)\bigr ]\Bigr\} \quad , 
\end{eqnarray}
with the definition
\begin{equation}
H(f;a,b,c)=f^{1+a}\, _2F_1\bigl
(1;\frac{1+a}{b};\frac{1+a+b}{c};-f^b c\bigr) \quad ,
\end{equation}
where  $_2F_1$  is  the  hypergeometric function.   In  this  case  the solution
presents a crossover between two power law regimes. By direct examination of the
right  hand  side  of  equation  (\ref{diff}),  it  can  be  seen  that    where
$(\lambda/\mu-1)^{(-1/(q-r))}>>f$ the  solution takes  the form  $f(s) \sim  ((q
-1)\lambda  s)^{(-1/(q-1))}$  and  where  $(\lambda/\mu-1)^{(-1/(q-r))}<<f$   it
becomes $f(s) \sim ((r-1)\lambda s)^{(-1/(r-1))}$.
\end{itemize}

In all the above  discussion the constant of  integration was chosen in  such a
way that $f(0)=1$, however in  order to introduce a different  normalisation it
is possible to include a normalising constant  $A$ as a change of scale in  the
dependent variable, $f \to f/A$.

The  analytical expression  for the  rank distribution  in the  general case  ,
equation (\ref{tsallis3}),  is rather  cumbersome to  work with.  However, it is
possible to derive a much simpler relation for the probability density  function
$p_f(f)$.  The value  of the  rank for  a word  with a  normalised frequency  of
occurrence $f$, can be written in the following way :
\begin{equation}
\label{prob}
s(f)=\int_f^{\infty}N p_f'(f')\, d f' \quad .
\end{equation}
This can  be seen  by noting  that $N  p_f'(f')$ gives  the number of words that
appear with normalised  frequency $f'$, thus  the corresponding position  in the
rank list of  a word with  frequency $f$ equals  the total number  of words that
have frequency greater or equal than $f$.  In addition, we can also write:
\begin{equation}
\label{basic}
s(f)=\int_f^{\infty}-\frac{d s}{d f'}\, d f' \quad .
\end{equation}
Since expressions (\ref{prob}) and (\ref{basic}) hold for any value of $f$,  the
following relation can be established:
\begin{equation}
p_f(f) \propto - \frac{d s}{d f} \quad .
\end{equation}
Moreover, the proportionality with the probability density as a function of
$n= N f$ is straightforward since $p_f(f)=N p_n(n)$. Finally, all these
considerations allow us to write the following  relation between the probability
densities and equation \ref{diff}):
\begin{equation}
\label{density}
p_n(n)=\frac{1}{N} p_f(f)\propto \frac{1}{\mu f^r+(\lambda-\mu)f^q} \quad \,
\end{equation}
This result is paticularly interesting in view of the mathematical
simplicity of equation (\ref{density}). Thereby, in essence we can interpret
 differential equation (\ref{diff}) as a model for the
functional form of $p_n(n)$.

Now we  proceed to  test the  phenomenological scenario  we have  just deployed
against empirical data from actual text sources. In Figures~4 and 5 we can  see
the  fit obtained  with equation  (\ref{tsallis2}) for  two of  the large  text
samples already used in Figure~2.  Whereas Zipf-Mandelbrot law would have  only
fitted a  small percentage  of the  total vocabulary  present in these corpora,
equation  (\ref{tsallis2})  captures   the  behaviour  of   the  frequency-rank
distribution along the whole range of the rank variable.

\begin{figure}
\label{figure4}
\epsfig{file=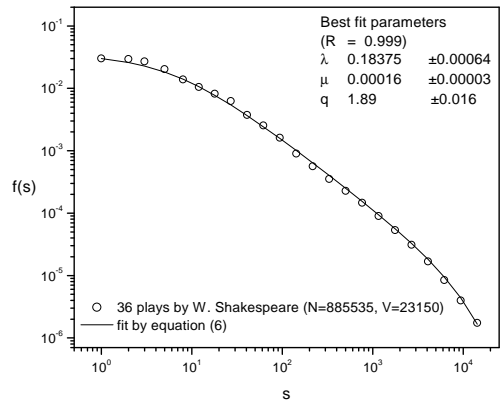,width=83mm}
\caption{Frequency-rank distribution for a corpus made up of 36 plays by William
Shakespeare   (circles)   together  with   a   fit  (full   line)   by  equation
(\ref{tsallis2}).}
\end{figure}

In  the  case  of   the  text  corpus  of   2606  books  the  general   solution
(\ref{tsallis3}) must be used in  order to fit the frequency-rank  distribution.
Alternatively,  by  performing  a normalised  histogram  of  the frequencies  of
appearance  of each  word, we  can recast  the data  in a  form suitable  to be
described by  equation (\ref{density}).  Figure~6 shows  the probability density
function $p_n(n)$ for the corpus of  2606 books, together with the fit  obtained
with equation (\ref{density}),  rendering again a  very good agreement  with the
actual  data.  Figure~7 depicts  the  frequency-rank distribution  for  the same
corpus compared  with a  plot of  equation (\ref{tsallis3})  using the  best fit
parameters presented in Figure~6 . In this case the parameter that controls  the
second power law regime takes the value $r=1.32$, indicating an asymptotic decay
exponent  close to  $-3$. In  the Figure  the range  of the  rank variable  was
extended  in order to make evident the whole transient between the two power law
regimes. Notwithstanding the good fit along the whole set of data points, it  is
clear that  a larger  text corpus  would be  necessary in  order to  see a fully
developed power law decay in the highest ranks that could allow a direct measure
of the exponent free from finite size effects.

\begin{figure}
\label{figure5}
\epsfig{file=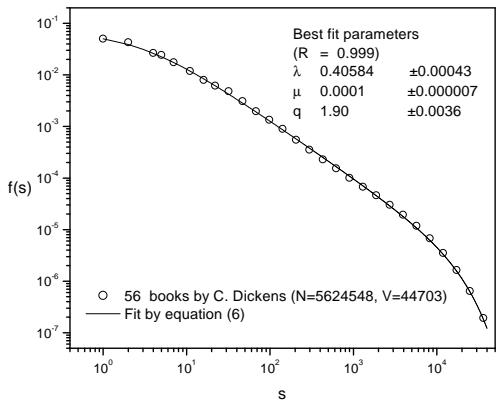,width=83mm}
\caption{Frequency-rank distribution for a corpus made up of 56 books by Charles
Dickens (circles) together with a fit (full line) by equation (\ref{tsallis2}).}
\end{figure}

\begin{figure}
\label{figure6}
\epsfig{file=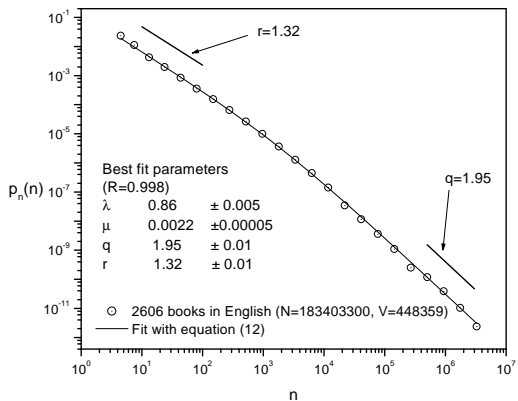,width=83mm}
\caption{Probability density function  $p_n(n) vs. n$  for  the large  corpus of
literary English and  the best fit  obtained with equation  (\ref{density}). The
straight lines show pure power laws that correspond to the asymptotic forms
of equation (12).}
\end{figure}

\begin{figure}
\label{figure7}
\epsfig{file=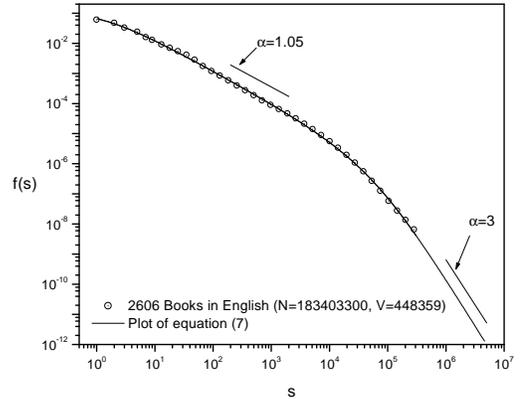,width=83mm}
\caption{Actual data from the  corpus of  2606 books in English together with  a
plot  of equation (\ref{tsallis3}) with the same paremeters shown in Figure~6. A
proportinality  constant  was   added  in   the  fitting  in  order  to   adjust
normalisation.}
\end{figure}

\section{Summary and conclusions}

The statistical evidence we  have presented in this  work has shown that  beyond
the  universal  features described  in  the narrow  range  of validity  of  Zipf
-Mandelbort law, lies  a vast region  spanning along the  rank axes where  a non
trivial macroscopic behaviour emerges when large text corpora are considered. We
have noticed that the majority of the words fall in the {\em non Zipfian} regime
showing a systematic  and robust statistical  behaviour. We have  also discussed
the natural  division of  words into  two different  kind of  vocabularies, each
prone to distinct linguistic usage. For large text samples with a high degree of
homogeneity we found that words for which the value of their rank  $s<3000-4000$
obey Zip-Mandelbrot law regardless of text length. The rest of the words,  whose
number may differ  considerably, all fall  in the fast  decaying  tails that  we
recognised in Figure~2. A more  profound study of these features  could possibly
shed light  on  aspects of how  language is used  and processed by  individuals.
Stepping up  two orders  of magnitude  in text  size a  new and more interesting
behaviour was   noticed in  agreement with  previous studies.  The analysis of a
huge corpus shows the  collective use  of language by a society of  individuals,
and more complex features were indeed observed as confirmed by the second  power
law regime for words beyond  $s \approx 5000-10000$. Moreover, all  the variants
of the non  trivial  phenomenology we  have just discussed  could be encompassed
within a  single mathematical  framework that  accurately accounts  for all  the
observed  features. After  the evidence  supplied by  this work  it seems  quite
plausible that  there may  be a  deep connection  between differential  equation
(\ref{diff}) and  the actual  processes underlying  the generation  of syntactic
language .

\section{Acknowledgements}

I am deeply indebted to Constantino  Tsallis and Dami\'an Zanette for all  their
suggestions  and very insightful observations.  I also thank  F. A.  Tamarit, S.
A.  Cannas  and  P. Pury  for valuable  discussions after  reading the  original
manuscript.

The text files analysed  in this paper were  supplied by {\em Project  Gutenberg
Etext} \cite{gut}.

This work was partially supported by the Secretary of Science and Technology of
the Universidad Nacional de C\'ordoba, Argentina.

\end{multicols}
\end{document}